\documentclass[aps,prl,twocolumn,showpacs,groupedaddress,floatfix]{revtex4}  
\usepackage{times}
\usepackage{graphicx}  
\usepackage{epsfig}
\usepackage{dcolumn}   
\usepackage{bm}        
\usepackage{amssymb}   
\usepackage{xspace}

\begin{document}
\newcommand{\dzero} {D0\xspace} \newcommand{\ttbar}
{\ensuremath{t\bar{t}}\xspace} \newcommand{\ppbar}
{\ensuremath{p\bar{p}}\xspace} \newcommand{\thetad}
{\ensuremath{\theta^{*}}\xspace}
\newcommand{\costheta}{\ensuremath{\cos\thetad}\xspace}
\newcommand{\ljets}{\ensuremath{\ell +}jets\xspace}
\newcommand{\mujets}{\ensuremath{\mu+jets}\xspace}
\newcommand{\ejets}{\ensuremath{e+jets}\xspace}
\newcommand{\fplus}{\ensuremath{f_{+}}\xspace}
\newcommand{\fminus}{\ensuremath{f_{-}}\xspace}
\newcommand{\fzero}{\ensuremath{f_{0}}\xspace}
\newcommand{\wjets}{\ensuremath{W+}jets\xspace} 
\newcommand{\met}{\mbox{$\not\!\!E_T$}\xspace} 
\newcommand{\tld}{\ensuremath{{\cal D}}\xspace} 
\newcommand{\tWb}{\ensuremath{t \rightarrow Wb}\xspace} 
\newcommand{\SM}{standard model\xspace} 
\newcommand{\MC}{Monte Carlo\xspace} \newcommand{\CL}{C.L.\xspace}
\newcommand{\deteta}{\ensuremath{\eta_{\mathrm{det}}}}
\newcommand{\alpgen}{{\sc{alpgen}}\xspace}
\newcommand{\pythia}{{\sc{pythia}}\xspace}
\newcommand{\geant}{{\sc{geant3}}\xspace}
\hyphenation{an-aly-sis}





\hspace{5.2in} \mbox{FERMILAB-PUB-07/588-E}

\title{Model-independent measurement of the {\boldmath W} boson helicity in top quark decays at D0}


%
\author{V.M.~Abazov$^{36}$}
\author{B.~Abbott$^{76}$}
\author{M.~Abolins$^{66}$}
\author{B.S.~Acharya$^{29}$}
\author{M.~Adams$^{52}$}
\author{T.~Adams$^{50}$}
\author{E.~Aguilo$^{6}$}
\author{S.H.~Ahn$^{31}$}
\author{M.~Ahsan$^{60}$}
\author{G.D.~Alexeev$^{36}$}
% ...

\affiliation{\vspace{0.1 in}(The D\O\ Collaboration)\vspace{0.1 in}}
% ...
\affiliation{$^{6}$University of Alberta, Edmonton, Alberta, Canada,
                Simon Fraser University, Burnaby, British Columbia, Canada,
                York University, Toronto, Ontario, Canada, and
                McGill University, Montreal, Quebec, Canada}
% ...
\affiliation{$^{29}$Tata Institute of Fundamental Research, Mumbai, India}
% ...
\affiliation{$^{31}$Korea Detector Laboratory, Korea University, Seoul, Korea}
% ...
\affiliation{$^{36}$Joint Institute for Nuclear Research, Dubna, Russia}
% ...
\affiliation{$^{50}$Florida State University, Tallahassee, Florida 32306, USA}
% ...
\affiliation{$^{52}$University of Illinois at Chicago,
                Chicago, Illinois 60607, USA}
% ...
\affiliation{$^{60}$Kansas State University, Manhattan, Kansas 66506, USA}
% ...
\affiliation{$^{66}$Michigan State University,
                East Lansing, Michigan 48824, USA}
% ...
\affiliation{$^{76}$University of Oklahoma, Norman, Oklahoma 73019, USA}
% ...


%


\date{December 11, 2007}

\begin{abstract}


 We present the first model-independent measurement of the helicity of $W$ bosons produced in top quark decays, based on a $1$~fb$^{-1}$ sample of
candidate $t\bar{t}$ events in the dilepton and lepton plus jets channels 
collected by the D0
detector at the Fermilab Tevatron $p\bar{p}$ Collider.
We reconstruct the
angle $\theta^*$ between the momenta of the down-type fermion and the top quark in the $W$ boson
rest frame for each top quark decay.
A fit of the resulting \costheta\ distribution
finds  that the fraction of longitudinal $W$ bosons $f_0 = 0.425 \pm 0.166 \hbox{ (stat.)} \pm  0.102 \hbox{ (syst.)}$ and the fraction of right-handed $W$ bosons $f_+ = 0.119 \pm 0.090 \hbox{ (stat.)} \pm  0.053 \hbox{ (syst.)}$, which is consistent at the 30\% C.L. with the standard model.


\end{abstract}


\maketitle 



The top quark is by far the heaviest of the known fermions and is
the only one that has a Yukawa coupling  to the Higgs boson of order unity
in the standard model (SM).   
In the SM, the top quark decays via the $V-A$ charged-current interaction,
almost always 
to a $W$ boson and a $b$ quark.  We search for evidence of new physics
in the  \tWb decay
by measuring the helicity of the $W$ boson.
A different Lorentz structure of the \tWb interaction would alter the fractions of $W$ bosons produced in each polarization state from the SM values of 
$0.697\pm0.012$~\cite{fzero} and $3.6\times10^{-4}$ ~\cite{fischer1} for the longitudinal fraction
\fzero and right-handed fraction \fplus, respectively, at the world average top quark mass $m_t$ of 
$172.5 \pm 2.3~{\rm GeV}$~\cite{WAtopmass}.


In this Letter, we 
report a
simultaneous measurement of \fzero\ and \fplus (the negative helicity fraction \fminus\ is then fixed by the requirement that $f_- +  f_0 + f_+ = 1$).  This is the first model-independent $W$ boson helicity measurement.  
  A measurement of 
  the $W$ boson helicity fractions 
that  differs significantly from
the SM values 
would be an unambiguous indication of new physics~\cite{caoSUSY}-\cite{WangTC2}.
In addition, the 
model-independent $W$ boson helicity measurement can be combined with measurements of single
top production cross sections to fully specify the $tbW$ vertex~\cite{chen2005}.  

  Measurements of the $b \rightarrow s\gamma$ decay rate assuming the absence of gluonic penguin      
contributions have indirectly limited
the $V+A$ contribution in top quark decays to less than a few percent~\cite{sbg1}.
 Direct 
measurements of the longitudinal fraction (\fplus\ set to zero)
found $\fzero=0.85^{+0.16}_{-0.23}$~\cite{helicityCDF2007} and $\fzero=0.56\pm0.31$~\cite{helicityD0}. 
Direct measurements of \fplus ( \fzero\ set to 0.7) have found $\fplus = -0.02\pm0.08$ 
~\cite{helicityCDF2007-2} and $\fplus = 0.06 \pm 0.10$ ~\cite{helicityD02006}.  The analysis
presented here improves upon that reported in Ref.~\cite{helicityD02006}
by using a larger data set,  
employing enhanced event selection techniques, making use of hadronic $W$ boson decays, and introducing the model-independent analysis.

The angular distribution of the down-type decay products of the $W$ boson
(charged lepton or $d$, $s$ quark)
in the rest frame of the $W$ boson can be described by 
introducing the decay angle \thetad of the down-type fermion 
with respect to the top quark direction.
The dependence of the distribution of
$\cos\thetad$ on the $W$ boson helicity fractions,
\begin{equation}
\omega(c) \propto 2(1-c^2)\fzero+
(1-c)^2\fminus\
                +(1+c)^2\fplus,
\label{eq:omega}
\end{equation}
where
$c  = \cos\thetad$, forms the basis for our measurement.
We proceed by selecting a data sample
enriched in \ttbar events, reconstructing the four vectors of the two
top quarks and their decay products,
 and then
calculating \costheta. 
The down-type fermions in leptonic $W$ boson decays are the charged leptons. 
 For hadronic $W$ boson decays,
we choose a $W$ boson daughter jet at random to calculate $\cos\thetad$.  Since this introduces a sign ambiguity into the calculation, we consider only $| \cos\thetad | $ for 
hadronic $W$ boson decays.
The $| \costheta |$ variable does    
not discriminate between left- and right-handed $W$ bosons, but adds information for 
determining the fraction of longitudinal  $W$ bosons.
These distributions in \costheta are compared with 
templates for different $W$ boson helicity models, accounting for background and 
reconstruction
effects, using a binned maximum
likelihood method.  



This measurement uses a data sample recorded with the D0
experiment~\cite{d0det} that corresponds to an integrated luminosity of about 
$1$~fb$^{-1}$
of \ppbar collisions at $\sqrt{s}=1.96$~TeV.  
Events were selected by
the trigger system 
based on the presence of energetic leptons or jets.
The data sample consists of \ttbar candidate
events from the lepton plus jets (\ljets) decay channel 
$t\bar{t}\rightarrow W^+W^-b\bar{b}\rightarrow \ell\nu qq^{\prime}b\bar{b}$
 and the dilepton channel
$t\bar{t}\rightarrow W^+W^-b\bar{b}\rightarrow \ell\nu\ell^{\prime}\nu^{\prime}b\bar{b}$, where $\ell$ and $\ell^{\prime}$ are electrons or muons.  The \ljets final state is
characterized by one charged lepton, at least four jets, and large missing transverse energy (\met).  The
dilepton final state is characterized by two charged leptons,
at least two jets, and large \met.  In both final states, at least two of the jets are $b$ jets.

The \ljets event selection~\cite{xsec_topo} requires an isolated lepton  with
transverse momentum 
$p_T>20$~$\mathrm{GeV}$, no other lepton with $p_T>15$~$\mathrm{GeV}$
in the event, $\met>20$~GeV, and at least four jets.  In the dilepton channel, 
events are required to have two leptons with opposite
charge and $p_T > 15$ GeV and two or more jets.
Electrons are
required to have pseudorapidity~\cite{def_eta} $|\eta|<1.1$ in the \ljets channel and $|\eta|<1.1$ or $1.5 < |\eta| < 2.5$ in the dilepton
channel, and are identified by
their energy deposition, isolation, and shower shape in the calorimeter, and
information from the tracking system~\cite{xsec_topo}. 
Muons are identified using information from the muon and tracking 
systems and must be isolated from jets, significant calorimeter energy, and energetic tracks. 
They are required to have $|\eta|<2.0$.
Jets are reconstructed using a cone algorithm with cone 
radius 0.5~\cite{jetreco} and are required to have rapidity  $|y|<2.5$ and $p_T>20$~$\mathrm{GeV}$.
The \met is calculated from the vector sum of calorimeter cell energies, corrected to account for the
response of the calorimeter to jets and electrons and also for the momenta of identified muons.

We simulate \ttbar signal events with $m_t = 172.5$ GeV  
for different
values of $f_+$
with the \alpgen Monte Carlo (MC) program~\cite{alpgen} for the 
parton-level process (leading order) and \pythia~\cite{pythia} for 
gluon radiation and subsequent
hadronization.  We generate samples corresponding to each of the three $W$ boson helicity configurations
by reweighting the generated \costheta distributions.
 
Backgrounds in the \ljets channel arise mainly from \wjets production 
and multijet
production.
In the dilepton channel, backgrounds arise from processes such as $WW+$jets or $Z+$jets.   The MC samples used to model background events with real leptons
are also generated
using \alpgen and \pythia.   Both the signal and background MC samples are passed through a 
\geant~\cite{geant} simulation of the detector response and reconstructed with the same algorithms used for data.  In the \ljets channel we estimate the number  $N_{\rm mj}$ of multijet
background events from data, using the technique described
in Ref.~\cite{xsec_topo}.
We calculate $N_{\rm mj}$ for each bin in the
\costheta distribution from the data sample to obtain the
multijet \costheta templates.

To increase the signal purity,
a multivariate likelihood discriminant $\tld$~\cite{xsec_topo} with values in the range 0 to 1 is calculated using input 
variables which exploit
differences in kinematics and jet flavor.  The kinematic variables 
considered are:
$H_T$ (scalar sum of the
jet $p_T$ values), centrality ${\cal C}$ (the ratio of $H_T$ to the sum of the jet energies),  $k_{T{\rm min}}^\prime$ (the distance in $\eta-\phi$ space between the closest pair of jets multiplied by the $E_T$ of the lowest-$E_T$ 
jet in the pair and divided by the $E_T$ of the $W$ boson), the sum of all jet and charged lepton energies $h$,  the minimum dijet mass of the jet pairs $m_{jj{\rm min}}$,
aplanarity ${\cal A}$, sphericity ${\cal S}$~\cite{apla}, 
 \met, and the dilepton invariant mass
$m_{\ell\ell}$.  In the dimuon channel, the $\chi^2$ of a kinematic fit to the
$Z \rightarrow \mu\mu$ hypothesis $\chi^2_Z$~\cite{dilepxsec} is used instead of \met.

Since  jets in background events arise mostly from light quarks or
gluons while two of the jets in \ttbar events arise from $b$ quarks,  
we form a neural network discriminant between $b$ and light jets~\cite{NNb} with output value 
{\it NN}$_b$ that tends towards one for $b$ jets and towards zero for light jets..  
In the $\ell+$jets channels we use the average of the two largest 
{\it NN}$_b$ values to form a continuous variable $\langle ${\it NN}$_b \rangle$ whose value 
tends to be large for
\ttbar events and small for backgrounds, while in the dilepton channels the
{\it NN}$_b$ values for the two leading jets ({\it NN}$_{b_1}$, {\it NN}$_{b_2}$) are taken as separate variables. 

\begin{figure}
  \epsfig{file=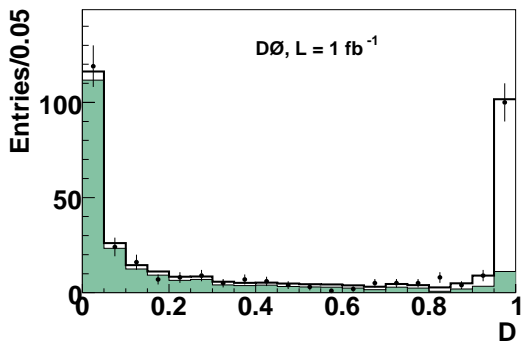,width=2.7in}
\caption{\label{fig:D} Distribution of ${\cal D}$ for data (points with error bars), background (shaded histogram), and signal plus background (open histogram) in the $e+$jets channel. }
\end{figure}

The discriminant is built separately for each of the five final states considered, using the method described in Refs.~\cite{xsec_topo,run1_topmass}. 
%
%
%
We consider all possible combinations of the above variables for use in the 
discriminant, and all possible requirements on the \tld value,
 and choose the variables and \tld criterion that give the best expected
precision for the $W$ boson helicity.   The variables chosen and the requirement 
placed on \tld for each channel are given in Table~\ref{tab:selection}.  An example of
the distributions of signal, background and data events in ${\cal D}$ is shown in 
Fig.~\ref{fig:D}.

We perform a binned Poisson maximum likelihood fit to compare
the observed distribution of events in \tld  
 to the sum of the
distributions expected from \ttbar and background  events.
In the $\ell+$jets channels, $N_{\rm mj}$ 
is constrained to the expected value within the known uncertainty, while in 
the dilepton channels the ratio of the various background sources is fixed
to the expectation from the cross sections times efficiency of the kinematic selection.
The likelihood is then
maximized with respect to the numbers of \ttbar and background events,  
which are multiplied by the efficiency for the \tld selection 
to determine the composition
of the sample used for measuring the $W$ boson helicity fractions.
Table~\ref{tab:selection} lists the composition of
each sample as well as the number of observed events in the data. 

 \begin{table*}
\caption{\label{tab:selection}%
Summary of the multivariate selection and number of selected events for each of the \ttbar\ final states used in this analysis.  The uncertainties are statistical only, except for the  background estimates 
in the $ee$ and $\mu\mu$ channels, in which systematic uncertainties arising from imperfections in the MC model of the data are included.
}
\begin{tabular}{lccccc}
\hline \hline
       &  $e+$jets & $\mu+$jets & $e\mu$ & $ee$ & $\mu\mu$  \\\hline
Variables used  in  & ${\cal C},$ $ {\cal S}$,  ${\cal A}$, $H_T$, & ${\cal C},$ $ {\cal S}$,  $H_T$,   &${\cal C},$ $ {\cal S}$, $h$, $m_{jj{\rm min}}$,       & ${\cal A}$,   $ {\cal S}$ , $k_{T{\rm min}}^\prime$, &    ${\cal A}$,   $ {\cal S}$ ,   $h$, $m_{jj{\rm min}}$,\\
     discriminant \tld       &   $h$,  $k_{T{\rm min}}^\prime$,  $\langle ${\it NN}$_b \rangle$     &   $k_{T{\rm min}}^\prime$,  $\langle ${\it NN}$_b \rangle$   &    $k_{T{\rm min}}^\prime$, {\it NN}$_{b_1}$, {\it NN}$_{b_2}$   &    \met, $m_{\ell\ell}$, {\it NN}$_{b_1}$      &    $\chi^2_Z$,    $m_{\ell\ell}$, {\it NN}$_{b_1}$   \\ [0.05 in]
Signal purity before \tld selection   & 0.38 $\pm$ 0.04   & 0.44 $\pm$ 0.04  & $0.67  \pm 0.11$ &   $0.014 \pm 0.004$  &      $0.024 \pm 0.006$        \\
Requirement on \tld & $>0.80$ & $>0.40$ & $>0.08$ & $>0.986$ & $>0.990$     \\
Background after \tld selection     &    $21.1 \pm 4.5$    &  $33.0 \pm  5.2$ & $9.9 \pm  2.5$ & $2.2 \pm 0.9$ &      $4.8 \pm 3.4$   \\
Data events after \tld selection &   121    & 167    &   45  & 15    & 15 \\ \hline \hline
\end{tabular}
\end{table*}

\begin{figure}
\epsfig{file=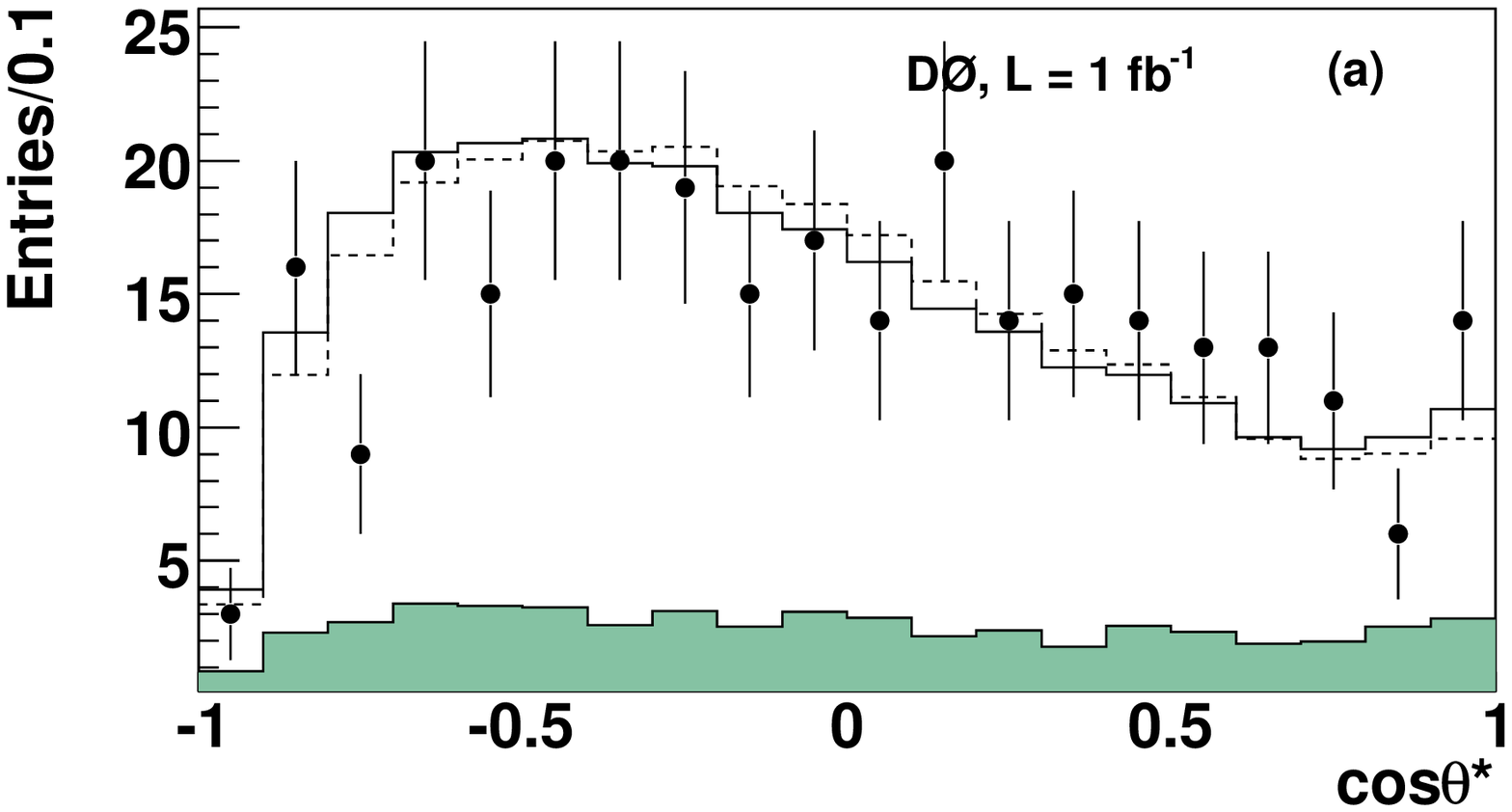, width=2.7in}
\epsfig{file=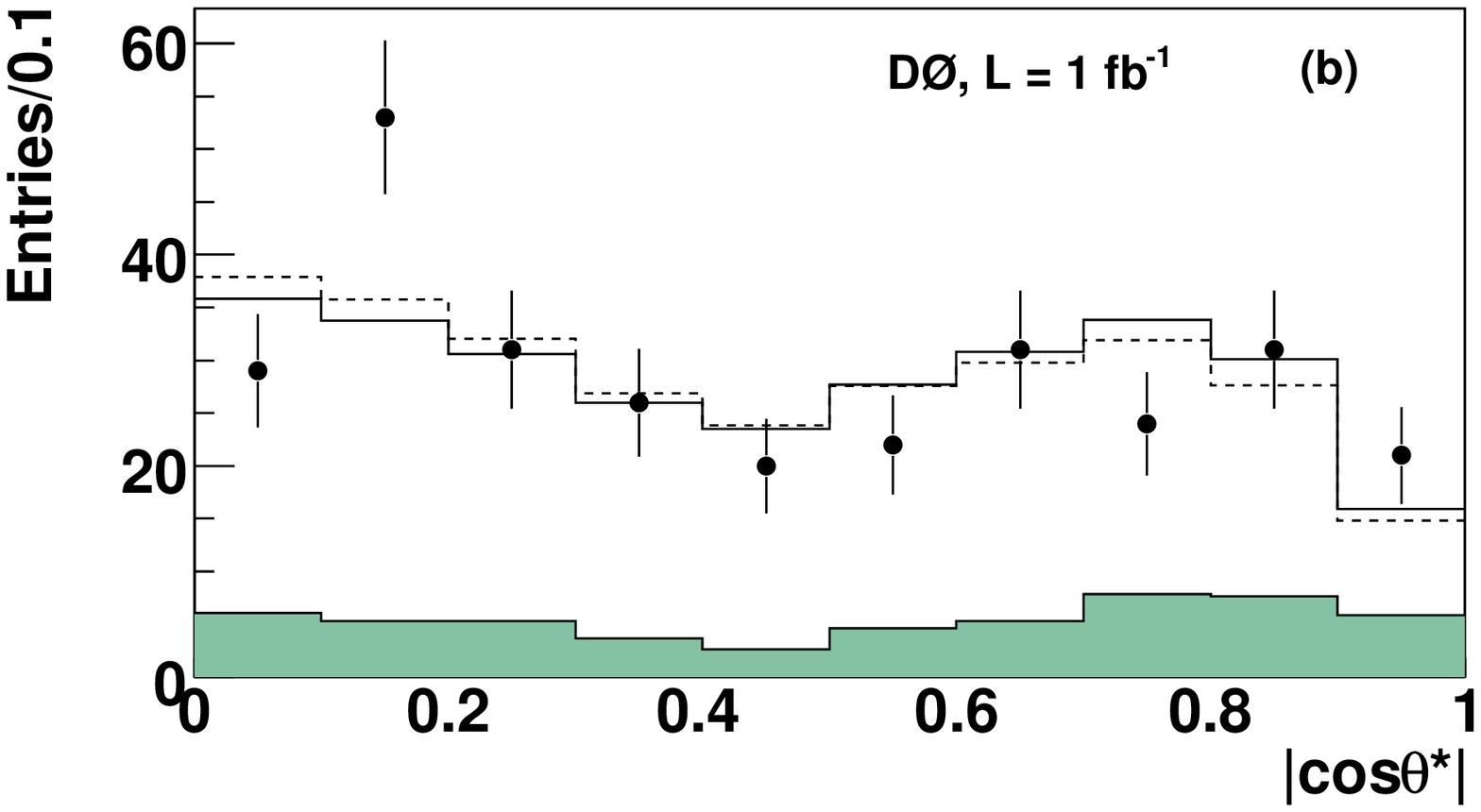, width=2.7in}
\epsfig{file=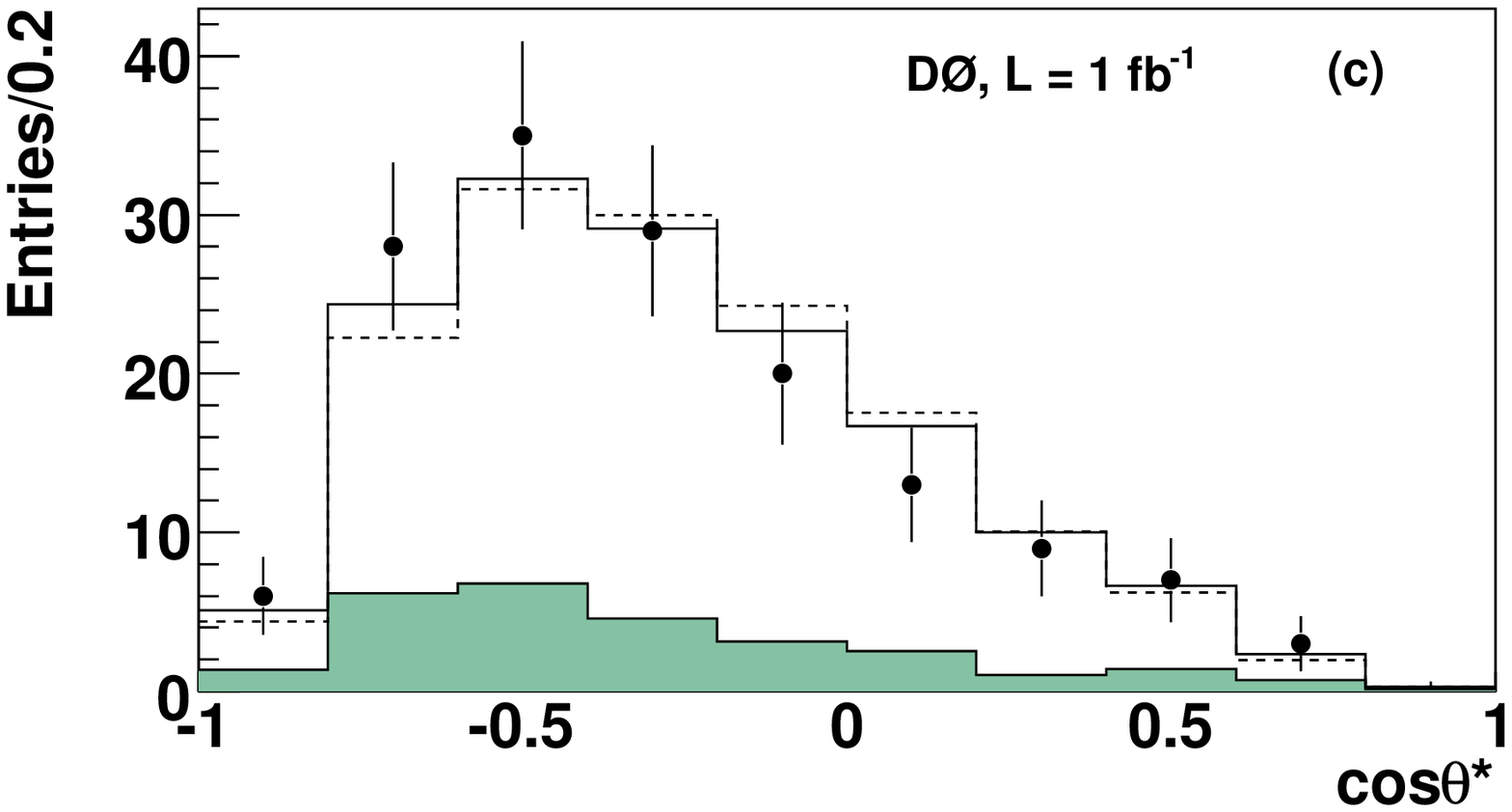, width=2.7in}
\caption{\label{fig:data2dmodel} 
Comparison of the $\cos\theta^*$ distribution in data (points with error bars) and the global best-fit model (solid open histograms) for (a) leptonic $W$ boson decays in \ljets\ events, (b) hadronic $W$ boson decays in \ljets\ events, and (c) dilepton events.  The dashed open histograms show the SM expectation, and the shaded histograms represent the background contribution.}
\end{figure}

The top quark and $W$ boson four-momenta in the selected \ljets events
are reconstructed using a
kinematic fit which is subject to the following constraints: two jets
must form the invariant mass of the $W$ boson~\cite{pdg}, the lepton and the \met
together with the neutrino $p_z$ component must form the invariant
mass of the $W$ boson, and the masses of the two reconstructed top quarks
must be $172.5$~$\mathrm{GeV}$.   The four highest-$p_T$ jets in each event are used
in the fit, and
among the twelve possible jet combinations, the solution with the 
maximal probability, considering both the $\chi^2$ from the kinematic fit
and the {\it NN}$_b$ values of the four jets, is chosen.
The
\costheta distributions for leptonic and hadronic $W$ boson decays 
obtained in the \ljets data after the full selection 
are shown in Fig.~\ref{fig:data2dmodel}(a) and (b).

Since the two neutrinos in the dilepton final state are not detected,  the system is kinematically underconstrained.
However, if $m_t$ is assumed, the 
kinematics
can be solved algebraically with a four-fold ambiguity in addition to the 
two-fold ambiguity in pairing jets with leptons.  For each of the two leading jets, we 
calculate the
value of \costheta resulting from each solution with each of the two leptons 
associated with the jet.  To explore the phase space consistent with the measured jet and lepton         
 energies, we fluctuate them according to their resolution                      
  many times, and repeat the above procedure for each fluctuation. 
The average of these values is taken as  \costheta for that jet.  The
\costheta distribution obtained in dilepton data is
shown in Fig.~\ref{fig:data2dmodel}(c). 

To extract $f_0$ and $f_+$, 
we  compute the binned Poisson likelihood $L(\fzero, \fplus)$ for the data to be consistent
with the sum of signal and background templates at
any given value for these fractions.  The background normalization is constrained to be
consistent within uncertainties with the expected value by a Gaussian term in the likelihood.
 The fit also accounts for the differences in selection efficiency for \ttbar\ events with different $W$ helicity
configurations~\cite{talk_with_doug}.


%
%

Systematic uncertainties 
are evaluated in ensemble tests
by varying the parameters 
that can affect the measurement.
Ensembles are formed by drawing events from a model with the parameter 
under study varied.  These are compared to the standard \costheta templates in a maximum 
likelihood fit.  The average shifts in the resulting \fzero 
and \fplus 
values are taken 
as the systematic uncertainty and are shown in Table~\ref{tab_systematic_2d}.
The total systematic uncertainty is then
taken into account in the likelihood by convoluting the likelihood with a 
Gaussian with a width that corresponds
to the total systematic uncertainty.
The mass of the top quark is
 varied by $\pm2.3$~$\mathrm{GeV}$,
and the jet reconstruction efficiency, energy calibration, 
and $b$ fragmentation
parameters by $\pm1\sigma$ around 
their nominal values.  
The $t\bar{t}$ model uncertainty is studied by
comparing 
$t\bar{t}$ events generated by \pythia\ to the standard  \alpgen\ samples, considering
samples with a different model for the underlying event and ones in which only a single primary vertex
is reconstructed.
%
Effects of mis-modeling the background distribution in \costheta\ are assessed by comparing
data to the background model for events with low \tld\ values.
 The uncertainty due to template 
statistics is evaluated by fluctuating the templates according to their statistical uncertainties and repeating the fit to the data for each fluctuation.  
Uncertainties due to jet resolution, jet flavor composition in the background, the modeling of the
{\it NN}$_b$ variable, and parton distribution functions are all found to be less than 0.01 for
both \fzero and \fplus.
\begin{table}[hhh]
\caption{\label{tab_systematic_2d} Summary of the major systematic uncertainties on $f_0$ and $f_{+}$
in the model-independent fit.}
\begin{tabular}{lcc}
\hline
\hline
Source & Uncertainty ($f_0$) & Uncertainty ($f_+$)  \\ \hline
Top mass               & 0.009 & 0.018 \\
Jet reconstruction eff.                & 0.021 & 0.010 \\
Jet energy calibration       & 0.012  &  0.019\\
$b$ fragmentation      & 0.016 &  0.010 \\ 
\ttbar ~model          & 0.068 & 0.032 \\
Background model       & 0.049   & 0.016  \\
Template statistics   & 0.049 & 0.025 \\ \hline
Total                  & 0.102  &   0.053 \\ \hline
 \hline
\end{tabular}
\end{table}



\begin{figure}
\includegraphics[trim=0 15 0 7,scale=0.40]{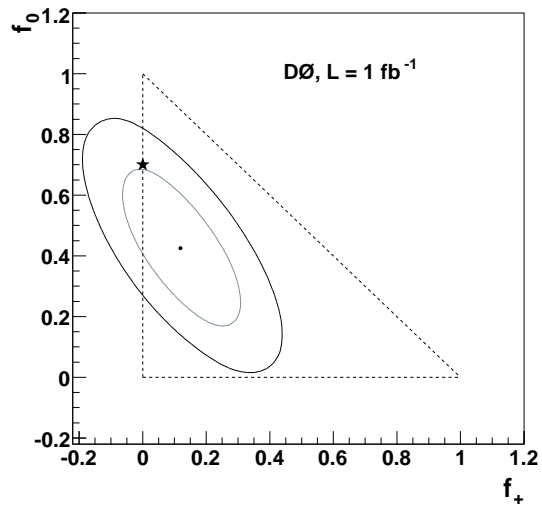} \\ 
\caption{\label{fig:data2dfit} Result of the model-independent $W$ boson helicity fit.  The ellipses are the 68\% and 95\% C.L. contours,  the triangle borders the physically allowed region where $f_0$ and 
$f_+$ sum to one or less, and the star denotes the SM values.}
\end{figure}

The measured values of $f_0$ and $f_+$ are:
\begin{eqnarray}
f_0 &=& 0.425 \pm 0.166 \hbox{ (stat.)} \pm  0.102 \hbox{ (syst.)} \\
f_+ &=& 0.119 \pm 0.090 \hbox{ (stat.)} \pm  0.053 \hbox{ (syst.)},  \nonumber
\end{eqnarray}
with a correlation coefficient of $-0.83$.
The inclusion of the $| \cos \theta^*|$ measurement from hadronic $W$ boson decays improves the     
uncertainties on $f_0$ and $f_+$ by about 20\% relative to those obtained using only the                                        
leptonic decays.  The 68\%, and 95\% C.L. contours from the  fit, including systematic uncertainties, are shown in Fig.~\ref{fig:data2dfit}.  The data indicate fewer longitudinal and more
right-handed $W$ bosons than the SM predicts, but the difference is not
significant as there is  a 30\% chance of observing a larger discrepancy given the statistical and systematic uncertainties in the measurement.  

If we fix $f_+$ to the SM value, we find
\begin{equation}
\label{eqf0_fixed}
f_0 = 0.619 \pm 0.090  \hbox{ (stat.)}  \pm 0.052  \hbox{ (syst.)},
\end{equation}
and if $f_0$ is fixed to the SM value we find
\begin{equation}
\label{eqfp_fixed}
f_+ = -0.002 \pm 0.047  \hbox{ (stat.)}  \pm 0.047  \hbox{ (syst.)}.
\end{equation}
Eqs.~\ref{eqf0_fixed} and~\ref{eqfp_fixed} are directly comparable to previous measurements~\cite{helicityCDF2007}-\cite{helicityD02006}.

In summary, we have measured the helicity fractions of $W$ bosons 
in \ttbar decays
in the \ljets and dilepton channels with a model-independent fit and find 
$f_0 = 0.425 \pm 0.166 \hbox{ (stat.)} \pm  0.102 \hbox{ (syst.)}$ and
$f_+ = 0.119 \pm 0.090 \hbox{ (stat.)} \pm  0.053 \hbox{ (syst.)}.$
 This is the first such measurement reported and is consistent at the 30\% level with  the SM values of $f_0 = 0.697$ and $f_+=3.6\times10^{-4}$.  We have also measured \fzero and \fplus in a model-dependent fit and find that they are consistent with the SM values.

%
We thank the staffs at Fermilab and collaborating institutions, 
and acknowledge support from the 
DOE and NSF (USA);
CEA and CNRS/IN2P3 (France);
FASI, Rosatom and RFBR (Russia);
CAPES, CNPq, FAPERJ, FAPESP and FUNDUNESP (Brazil);
DAE and DST (India);
Colciencias (Colombia);
CONACyT (Mexico);
KRF and KOSEF (Korea);
CONICET and UBACyT (Argentina);
FOM (The Netherlands);
Science and Technology Facilities Council (United Kingdom);
MSMT and GACR (Czech Republic);
CRC Program, CFI, NSERC and WestGrid Project (Canada);
BMBF and DFG (Germany);
SFI (Ireland);
The Swedish Research Council (Sweden);
CAS and CNSF (China);
and the Alexander von Humboldt Foundation.
%


\begin{thebibliography}{99}
\bibitem{def_eta} Rapidity $y$ and pseudorapidity $\eta$ are defined
  as functions of the polar angle $\theta$ with respect to the proton
  beam and the parameter $\beta$ as $y(\theta,\beta)\equiv
  \frac{1}{2}\ln[(1+\beta\cos\theta)/(1-\beta\cos\theta)]$ and
  $\eta(\theta)\equiv y(\theta,1)$, where $\beta$ is the ratio of a
  particle's momentum to its energy.
\end{thebibliography}
\end{document}